\documentclass[12pt]{article}
\usepackage[english]{babel}
\usepackage{ulem}
\usepackage[utf8]{inputenc}
\usepackage[T1]{fontenc}

\def\hbj{\hat{\bj}}
\def\bX{\mathbf{X}}

\def\bj{\mathbf{j}}
\def\hj{\hat{j}}

\def\hg{\hat{g}}

\usepackage{amsmath}

\def\mC{\mathcal{C}}
\usepackage{amsfonts}

\usepackage{url}
\usepackage[dvips]{graphicx}
\usepackage{calc}
\usepackage{subfigure}
\usepackage{multirow}

\def\mL{\mathcal{L}}
\def\mH{\mathcal{H}}

\def\bA{\mathbf{A}}

\def\hnabla{\hat{\nabla}}

\newcommand{\tg}{\tilde{g}}

\def\bD{\mathbf{D}}

\def\tj{\tilde{j}}

\begin{document}
	\begin{titlepage}
	\begin{center}
		{\Large{ \bf Ideal Fluids  In Born-Infeld Inspired Gravity }}
		
		\vspace{1em}  
		
		\vspace{1em} J. Kluso\v{n} 			
		\footnote{Email addresses:
			klu@physics.muni.cz (J.
			Kluso\v{n}) }\\
		\vspace{1em}
		\textit{Department of Theoretical Physics and
			Astrophysics, Faculty of Science,\\
			Masaryk University, Kotl\'a\v{r}sk\'a 2, 611 37, Brno, Czech Republic}
		
		\vskip 0.8cm
		
		%
		%
		%
		%
		%
		%
		
		\vskip 0.8cm
		
	\end{center}

\begin{abstract}
In this short note we study Born-Infeld Inspired Gravity together with an action functional for ideal fluid. We obtain corresponding equations of motion and also determine canonical form of this action. 
	\end{abstract}

\bigskip

\end{titlepage}

\newpage

\section{Introduction and Summary}\label{first}
Born-Infeld inspired modification of gravity (BIMG) is very interesting extension of standard Einstein-Hilbert action 
\cite{Banados:2010ix} which is based on the simple idea to apply structure of Born-Infeld action \cite{Leigh:1989jq}
for electrodynamics to the case of gravity
\footnote{For review and extensive list of references, see \cite{BeltranJimenez:2017doy}.}. Very important property of 
BIMG is that it has to be formulated in Palatini formalism where connection and metric are independent variables
since Born-Infeld gravity with metric as fundamental variable
contains higher orders of Ricci tensor in the action so 
that this theory contains derivatives of second order and hence
suffers with related instabilities as was explicitly demonstrated in 
\cite{Deser:1998rj}. Alternatively we can say that BIMG was inspired by Eddington's formulation of gravity 
where the only dynamical degrees of freedom are components of connection. This theory is equivalent to Einstein-Hilbert action in the absence of matter. In fact, it is very difficult to include matter into Eddington gravity, for nice discussion see \cite{Chakraborty:2020yag} and for recent proposal that works in case of scalar fields, see \cite{Kluson:2025dbc}. The reason why it is difficult to include matter is that action for matter generally depends on metric in non-trivial way and then the only possibility how to couple it to Eddington like gravity is BIMG action where now metric is non-dynamical field which provides coupling between matter sector and gravitational degrees of freedom. This fact is especially nicely seen when we perform canonical analysis of BIMG \cite{Kluson:2025dbc,Kluson:2025ajf,Kluson:2025uyl} where it was explicitly shown that  dynamical metric degrees of freedom emerge as momenta conjugate to connection.

Natural extension of this analysis is to study coupling of BIMG to currents. This is  important problem since matter currents are fundamental in cosmological application. It is clear that since current is vector as  a geometrical object with upper space-time indices it is not possible to include it directly into Eddington gravity since we have to lower its space-time index using metric. For that reason we should incorporate currents directly into BIMG actions. Then in order to describe current contribution we follow fundamental work 
\cite{Brown:1992kc} where an action for perfect fluid was proposed. Then we determine corresponding equations of motion. We explicitly show that the divergence of the stress energy tensor with respect to the covariant derivative 
compatible with non-dynamical metric is zero. We  also suggest relation between dynamical metric, non-dynamical one and current and we argue that gravitational equations of motion have formally the same form as in case of General Relativity when the matter contribution is given by the stress energy tensor with however complicated equation of state. We also propose an alternative form of BIMG action that couples to current but we show that after sequence of redefinitions of metric and current it reduces to the original BIMG action coupled minimally to perfect fluid.

As the final part of our analysis we determine canonical form of BIMG action coupled to perfect fluid. Since the gravitational part is the same as in case of BIMG action we use results derived recently in \cite{Kluson:2025dbc,Kluson:2025ajf,Kluson:2025uyl}
where canonical form of BIMG action was studied. In case of perfect fluid  we proceed in the similar way as in 
\cite{Brown:1992kc} and find canonical form of the action. Finally we integrate out non-dynamical metric. Resulting action is still given as a sum of four constraints which are however more complicated than in case of General Relativity.

This paper is organized as follows. In the next section (\ref{second}) we introduce BIMG coupled to the perfect fluid, derive corresponding equations of motion and study their general solutions. Then we also introduce more general form of BIMG action coupled to current and argue that after suitable chain of redefinition it  transforms to original BIMG action coupled to perfect fluid. Finally in section (\ref{third}) we determine 
canonical form of BMIG action coupled to perfect fluid.

\section{Born-Infeld Inspired Gravity Coupled with Matter Current}
\label{second}
In this section we introduce BIMG action  that couples minimally to matter current. Such an action has the form 
\begin{eqnarray}\label{act1}
	&&	S=S_{BIMG}+S_{matt} \ , \quad 
S_{BIMG}=M_p^2M_{BI}^2\int d^4x [\sqrt{-\det \bA}-\lambda\sqrt{-g}] \ , \nonumber 
	\\
	&&
	\bA_{\mu\nu}=g_{\mu\nu}+\frac{1}{M^2_{BI}}R_{(\mu\nu)} \  , 
	\nonumber \\
	&&  S_{matt}=\int d^4x \sqrt{-g}[F(\bj)+j^\mu\partial_\mu \phi]\ , \quad \bj=\sqrt{-j^\alpha
		g_{\alpha\beta}j^\beta} \ .  \nonumber \\
\end{eqnarray}
 $S_{BIMG}$ is standard BIMG  action 
where $g_{\mu\nu}$ is independent metric that couples to matter through the action $S_{matt}$. Further, $R_{(\mu\nu)}=
\frac{1}{2}(R_{\mu\nu}+R_{\nu\mu})$ is symmetrized Ricci tensor that is function  of independent connection $\Gamma^\rho_{\mu\nu}$ which we choose to be symmetric $\Gamma^\rho_{\mu\nu}=\Gamma^\rho_{\nu\mu}$. Finally $M_p$ is Planck mass  scale and $M_{BI}^2$ is the scale of  new physics when the General Relativity is recovered in the limit $M_{BI}\rightarrow \infty$. Finally $\lambda$ is non-zero dimensionless constant 
that is needed for consistency of Born-Infeld inspired gravity, 
for review and extensive list of references, see 
\cite{BeltranJimenez:2017doy}.

Let us discuss in more details the matter action $S_{matt}$. 
Such a form of action 
was firstly introduced in \cite{Brown:1992kc} and we mainly follow notation used in 
\cite{Ferreira:2020fma}. Function $F(\bj)$ that appears in $S_{matt}$ is an arbitrary function 
of $\bj$ where $j^\mu$ is vector current, 
 $\phi$ is  scalar field that is introduced to ensure that $j^\mu$ is conserved on-shell as we will explicitly show below. 

In order to  analyze 
Lagrangian equations of motion it is convenient to introduce 
an auxiliary metric $\hg_{\mu\nu}$ and its inverse $\hg^{\mu\nu}, 
\hg_{\mu\nu}\hg^{\nu\rho}=\delta_\mu^\rho$ and rewrite the action $S_{BIMG}$ that appears in 
(\ref{act1}) into the form 
\begin{eqnarray}\label{actaux}
	&&	
	S_{BMIG}=M_p^2M_{BI}^2\int d^4x \left[\frac{1}{2}\sqrt{-\det \hg}(\hg^{\mu\nu}\bA_{\mu\nu}-2)-\lambda\sqrt{-g}\right] \ . \nonumber 
	\\
\end{eqnarray}
To see an equivalence between (\ref{actaux}) and (\ref{act1}) let us consider equations of motion for $\hg^{\mu\nu}$
\begin{equation}
	-\frac{1}{2}\hg_{\mu\nu}(\hg^{\rho\sigma}\bA_{\rho\sigma}-2)+
	\bA_{\mu\nu}=0
\end{equation}
that has solution 
\begin{equation}
	\hg_{\mu\nu}=\bA_{\mu\nu} \ . 
\end{equation}
Inserting this result into action (\ref{actaux})  we easily get that it reduces to the original action 
(\ref{act1}).

Now we are ready to study equations of motion that follow from 
the action (\ref{actaux}). 
 Performing variation with respect to $j^\alpha,\phi$ we get following equations of motion for current and scalar field $\phi$
\begin{eqnarray}\label{eqcurrent}
	-\frac{dF}{d\bj}\frac{g_{\mu\nu}j^\nu}{\bj}+\partial_\mu \phi=0 \ , \quad 
	\partial_\alpha[\sqrt{-g}j^\alpha]=0 \ . \nonumber \\
\end{eqnarray}
Note also that the first equation in (\ref{eqcurrent}) implies
\begin{equation}\label{rel1}
	\nabla_\mu \left(\frac{dF}{d\bj}\frac{j_\nu}{\bj}\right)=
	\nabla_\nu \left(\frac{dF}{d\bj}\frac{j_\mu}{\bj}\right)  \ , \quad j_\mu=g_{\mu\nu}j^\nu
\end{equation}
as follows from the fact that $\nabla_\mu\partial_\nu\phi=
\nabla_\nu \partial_\mu\phi$, where we  introduced covariant derivative $\nabla_\mu$ that is compatible
with the metric $g_{\mu\nu}$ 
\begin{equation}\label{nablag}
	\nabla_\rho g_{\mu\nu}=0 \ . 
\end{equation}	
Note that with the help of covariant derivative $\nabla_\mu$ it is possible to rewrite
the second equation in (\ref{eqcurrent}) into the form 
\begin{equation}\label{nablaj}
\nabla_\mu j^\mu=0
\end{equation}
as follows from the fact that $\nabla_\mu V^\mu=\frac{1}{\sqrt{-g}}\partial_\mu[\sqrt{-g}V^\mu]$ for any vector field $V^\mu$. 

For further purposes we rewrite (\ref{rel1}) into the form 
\begin{equation}\label{rel2}
	\nabla_\mu\left(\frac{dF}{d\bj}\frac{j^\nu}{\bj}\right)=
g^{\nu\omega}\nabla_\omega\left(\frac{dF}{d\bj}\frac{j^\sigma}{\bj}\right)g_{\sigma\mu}
\end{equation}
that will be useful bellow. 
Finally we determine equations of motion for $\Gamma^\rho_{\mu\nu}$ and $g_{\mu\nu}$ from the action (\ref{actaux}). Let us firstly perform variation of the action (\ref{actaux})
 with respect to $\Gamma^\rho_{\mu\nu}$ and we obtain 
\begin{eqnarray}\label{parbD}
	-\partial_\rho \bD^{\mu\nu}+\bD^{\alpha\beta}\Gamma_{\alpha\beta}^\nu\delta^\mu_\rho
	+\bD^{\mu\nu}\Gamma^\sigma_{\sigma \rho}-\bD^{\mu \sigma}\Gamma^\nu_{\sigma\rho}-\bD^{\sigma \nu}
	\Gamma^\mu_{\sigma \rho}+
	\partial_\sigma[\bD^{\sigma \nu}\delta_\rho^\mu]=0 \ , \nonumber \\
\end{eqnarray}
where we introduced symmetric tensor density
\begin{equation}
	\bD^{\mu\nu}\equiv \sqrt{-\det\hg}\hg^{\mu\nu} \ . 
\end{equation}
Note that the equation (\ref{parbD})  can be rewritten with the help of covariant derivatives $\hnabla_\mu$ into the form
\begin{eqnarray}\label{eqmPal}
	-\hnabla_\rho \bD^{\mu\nu}+\hnabla_\sigma\bD^{\sigma \nu}\delta_\rho^\mu=0 \ , 
	\nonumber \\
\end{eqnarray}
where the covariant derivative of tensor density $\bD^{\mu\nu}$  of weight $-1$ is equal to 
\begin{equation}
	\hnabla_\rho \bD^{\mu\nu}=\partial_\rho \bD^{\mu\nu}+\bD^{\mu \sigma}\Gamma_{\sigma\rho}^\nu+
	\Gamma^\mu_{\rho \sigma}\bD^{\sigma \nu}-\bD^{\mu\nu}\Gamma^\sigma_{\sigma\rho} \ . 
\end{equation}
Further, performing contraction between $\mu$ and $\rho$ (\ref{eqmPal}) we get
\begin{equation}
	\hnabla_\rho \bD^{\rho\nu}=0 \ . 
\end{equation}
Inserting this result into (\ref{eqmPal}) 
we finally obtain
\begin{equation}
	\hnabla_\rho \bD^{\mu\nu}=0 \ .
\end{equation}
Since $\hg$ is non-singular matrix the equation above can be rewritten into the form 
\begin{equation}
	\hnabla_\rho \hg^{\mu\nu}=0 \ . 
\end{equation}
In other words we get an important result that says that coefficients of connection 
$\Gamma^\rho_{\mu\nu}$ are uniqually defined with the metric $\hg_{\mu\nu}$ as
\begin{equation}
	\Gamma^\rho_{\mu\nu}=\frac{1}{2}\hg^{\rho\sigma}(\partial_\mu \hg_{\sigma\nu}+
	\partial_\nu \hg_{\sigma\mu}-\partial_\sigma \hg_{\mu\nu}) \ . 
\end{equation}
Once again we should stress that there are two covariant derivatives $\nabla,\hnabla$, where the
first one is defined by the metric $g_{\mu\nu}$ while the second one by the metric $\hg_{\mu\nu}$. 

As the last step  we perform variation of the action (\ref{actaux}) with respect to  $g_{\mu\nu}$ and we get
\begin{eqnarray}\label{eqg}
	M^2_pM^2_{BI}[\sqrt{-\hg}\hg^{\mu\nu}-\lambda \sqrt{-g}g^{\mu\nu}]+
	\sqrt{-g}T^{\mu\nu}=0 \ , \nonumber \\
\end{eqnarray}
where we defined $T^{\mu\nu}$ as 
\begin{eqnarray}\label{Tmunu}
	T^{\mu\nu}=\frac{2}{\sqrt{-g}}\frac{\delta S_{matt}}{\delta g_{\mu\nu}}=g^{\mu\nu}[F+j^\alpha\partial_\alpha \phi]-\frac{dF}{d\bj}\frac{j^\mu j^\nu}{\bj} \ .  \nonumber \\
\end{eqnarray}
Let us use equations of motion for $j^\mu$ to replace $\partial_\mu\phi$ with $\frac{dF}{d\bj}\frac{g_{\mu\nu}j^\nu}{\bj}$ so that stress energy tensor (\ref{Tmunu}) has the form 
\begin{eqnarray}\label{stresseneten}
	T^{\mu\nu}=g^{\mu\nu}[F-\bj\frac{dF}{d\bj}]-\frac{dF}{d\bj}\frac{j^\mu j^\nu}{\bj} \ .  \nonumber \\
\end{eqnarray}
It is crucial to show that the stress energy tensor (\ref{stresseneten}) 
is conserved with respect to the metric $g_{\mu\nu}$.  To show this let us 
 determine divergence of stress energy tensor
\begin{eqnarray}\label{consT}
&&\nabla_\mu T^{\mu\nu}=g^{\mu\nu}\nabla_\mu F-g^{\mu\nu}\nabla_\mu[\bj\frac{dF}{d\bj}]-
\nabla_\mu[\frac{dF}{d\bj}\frac{j^\nu}{\bj}]j^\mu=\nonumber \\	
&&=g^{\mu\nu}\nabla_\mu F-g^{\mu\nu}\nabla_\mu[\bj\frac{dF}{d\bj}]
+g^{\nu\omega}\nabla_\omega[\frac{dF}{d\bj}\bj]-g^{\nu\mu}
\nabla_\mu F =0 \ , \nonumber \\
\end{eqnarray}
where we used (\ref{nablag}),(\ref{nablaj}) and also the fact that 
\begin{eqnarray}
\nabla_\mu[\frac{dF}{d\bj}\frac{j^\nu}{\bj}]j^\mu=
-g^{\nu\omega}\nabla_\omega[\frac{dF}{d\bj}\bj]+g^{\nu\mu}
\nabla_\mu F \  \nonumber \\
\end{eqnarray}
as follows from (\ref{rel2}). Note that (\ref{consT}) is very important result that
proves conservation of stress energy tensor on condition when the matter equations of motion
are obeyed. This is crucial difference with respect to the case of General relativity when 
the stress energy tensor is covariantly constant as a consequence of gravitational equations
of motion and the fact that Einstein tensor has zero divergence.
We would like also stress that after performing an identification 
\cite{Brown:1992kc,Ferreira:2020fma}
\begin{eqnarray}
	n=\bj \ , \quad  u^\mu=\frac{j^\mu}{n} \ , \quad  F-\bj\frac{dF}{d\bj}=p , \quad  \rho=-F
	\nonumber \\
\end{eqnarray}
we obtain stress energy tensor in the familiar form 
\begin{equation}
	T^{\mu\nu}=p g^{\mu\nu}+(\rho+p) u^\mu u^\nu \ 
\end{equation}
that corresponds to the stress energy tensor of ideal fluid.

Let us finally return to the equations of motion for $\hg$ and write them in an explicit form 
\begin{eqnarray}\label{eqR}
R_{\mu\nu}(\Gamma(\hg))-\frac{1}{2}R(\Gamma(\hg))\hg_{\mu\nu}=M^2_{BI}(g_{\mu\nu}-\hg_{\mu\nu}+
\frac{1}{2}\hg_{\mu\nu}(\hg^{\rho\sigma}g_{\rho\sigma}-2)) \ ,  \nonumber \\
\end{eqnarray}
where the left side has the standard form of the equations of motion for dynamical gravitational field
$\hg_{\mu\nu}$ while the right side depends on $g_{\mu\nu}$ and $\hg_{\mu\nu}$ where $g_{\mu\nu}$ can be expressed as function 
of $\hg_{\mu\nu}$ and matter from (\ref{eqg}). In fact, let us presume that this dependence can be written as
\begin{equation}\label{gans}
g^{\mu\nu}=X \hg^{\mu\nu}+Y j^\mu j^\nu \ , 
\end{equation}
where unknown functions $X$ and $Y$ can be determined from (\ref{eqg}). In more details, using an ansatz
(\ref{gans}) we obtain
\begin{equation}
	g_{\mu\nu}=\frac{1}{X}(\hg_{\mu\nu}-\frac{Y}{X}\frac{\hj_\mu \hj_\nu}{1-\frac{Y}{X}\hbj^2}) \ , 
	\hj_\mu=\hg_{\mu\nu}j^\nu \ , \quad \hbj^2=-j^\mu \hg_{\mu\nu} j^\nu \ . 
\end{equation}
Inserting this ansatz into the equation (\ref{eqg}) and comparing terms proportinal to $\hg^{\mu\nu}$ and $j^\mu j^\nu$ we obtain two equations for $X$ and $Y$ that can be solved for them at least in principle and let us denote these solutions as $X=\bX(\hbj,M^2_{BI}),Y=\mathbf{Y}(\hbj,M^2_{BI})$. Then  the equations of motion (\ref{eqR}) can be rewritten into the form 
\begin{eqnarray}
R_{\mu\nu}(\Gamma(\hg))-\frac{1}{2}R(\Gamma(\hg))\hg_{\mu\nu}=
\mathbf{X}(\hbj,M^2_{BI})\hg_{\mu\nu}+\mathbf{Y}(\hbj,M^2_{BI})
\hj_\mu \hj_\nu\equiv \hat{T}_{\mu\nu} \ , \nonumber \\
\end{eqnarray}
where $\hat{T}_{\mu\nu}$ is stress energy tensor for ideal fluid with complicated equations of state as follows from the structure of functions $\mathbf{X},\mathbf{Y}$. Note however that the new stress energy tensor $\hat{T}_{\mu\nu}$ is automatically conserved with respect to the covariant derivative $\hat{\nabla}$ which is a consequence of the identity 
\begin{equation}
	\hnabla_\mu (R^{\mu\nu}(\Gamma(\hg))-\frac{1}{2}\hg_{\mu\nu}
	\hat{R}(\Gamma(\hg)))=0 \ . 
\end{equation}
Generally it is very complicated to find dependence of $g_{\mu\nu}$ on $\hg_{\mu\nu}$ and $j^\mu$. However it is interesting that resulting equations of motion for $\hg$ still have
the same form as in General Relativity where all corrections coming from the DBI structure of gravitational action are included into the new stress energy tensor.
\subsection{Alternative Form of the Current Contribution}
In previous section we studied BIMG action  that couples
to perfect fluid represented  in  minimal way which means that the whole action is sum of BMIG action with matter action. We can ask the question whether 
it is possible to consider more general form of current contribution. Let us propose 
following modified BIMG action 
\begin{eqnarray}\label{SBIC}
&&	S_{BIMG}=
	M^2_{BI}M^2_p\int d^4x [\sqrt{-\det \tilde{\bA}}+(\tj^\alpha\partial_\alpha\phi-\lambda)\sqrt{-g}] \ , \nonumber \\ 
&&	\tilde{\bA}_{\mu\nu}=\tg_{\mu\nu}+\frac{1}{M^2_{BI}}\left(R_{(\mu\nu)}+\frac{1}{2M_p^2}\tilde{F}(\tilde{\bj})\right) \tg_{\mu\nu} \ , \quad \tilde{\bj}^2=-\hj^\mu \tg_{\mu\nu}\hj^\nu \ ,
\end{eqnarray}
where we introduced current contribution under the determinal structure while we still keep term $\tj^\alpha \partial_\alpha \phi\sqrt{-\tg}$ in order to ensure that current $\tj^\alpha$ is conserved on-shell. Further, it is easy to see that this action  reduces into General Relativity action in Palatini formulation coupled minimally to current in   the limit $M_{BI}\rightarrow \infty$ when we can write
\begin{equation}
	\sqrt{-\det \bA}=
	\sqrt{-\det \tg}(1+\frac{1}{2M^2_{BI}}\tg^{\mu\nu}R_{(\mu\nu)}(\Gamma)+
	\tilde{F}(\tilde{\bj})
\end{equation}
so that the action (\ref{SBIC}) is equal to 
\begin{eqnarray}
&&	S_{BIC}(M_{BI}\rightarrow \infty)=
	\int d^4x [\frac{1}{2}\sqrt{-\tg}\tg^{\mu\nu}R_{\mu\nu}(\Gamma)+M^2_{BI}(1-\lambda)\sqrt{-\tg}]+\nonumber \\
&&+	\int d^4x [\tilde{F}(\tilde{\bj})+\tj^\mu\partial_\mu\phi]
\end{eqnarray}
which is a Palatini aciton for General Relativity coupled to ideal fluid. 

Let us now study  (\ref{SBIC}) in more details. As the first step we perform following
redefinition of metric 
\begin{equation}
	\tg_{\mu\nu}+\frac{1}{2}\tg_{\mu\nu}\tilde{F}(\tilde{\bj}^2)=g_{\mu\nu}
\end{equation}
from which we also get
\begin{eqnarray}
	\tilde{\bj}^2(1+\frac{1}{2}\tilde{F}(\tilde{\bj}^2))=\tj^\mu g_{\mu\nu}\tj^\nu  \ 
	\nonumber \\
\end{eqnarray}
that can be solved for $\tilde{\bj}$ at least in principle
\begin{equation}
	\tilde{\bj}^2=\tilde{\bj}^2(\tj^\mu g_{\mu\nu}\tj^\nu) \ .
\end{equation}
After this redefinition the action (\ref{SBIC}) takes the form 
\begin{eqnarray}
&&	S_{BIMG}=
	M^2_{BI}M^2_p\int d^4x [\sqrt{-\det \bA}+(\tj^\alpha\partial_\alpha\phi-\lambda)\sqrt{-g}F'(\tj^\mu g_{\mu\nu}\tj^\nu)] \ , \nonumber \\ 
&&	\bA_{\mu\nu}=g_{\mu\nu}+\frac{1}{M^2_{BI}}R_{(\mu\nu)} \ , \quad 
	\nonumber \\
\end{eqnarray}
where we defined $F'(\tj^\mu g_{\mu\nu}\tj^\nu)$ as
\begin{equation}
	F'(\tj^\mu g_{\mu\nu}\tj^\nu))=\frac{1}
	{\sqrt{1+\frac{1}{2}\tilde{F}(\tilde{\bj}(
			\tj^\mu g_{\mu\nu}\tj^\nu))}} \ . 
\end{equation}
Finally we redefine $\tj^\alpha$ as
\begin{equation}
	\tj^\alpha F'(\tilde{\bj}(\tj^\mu g_{\mu\nu}\tj^\nu))=j^\alpha
\end{equation}
that implies
\begin{equation}
(	\tj^\mu g_{\mu\nu}\tj^\nu) F'^2(\tilde{\bj}(\tj^\mu g_{\mu\nu}\tj^\nu))=j^\alpha g_{\mu\nu}j^\beta
\end{equation}
so that we can again presume that this equation can be solved for $\tj^\mu g_{\mu\nu}\tj^\nu$ as function 
$\bj^2=-j^\mu g_{\mu\nu}j^\nu$. Then inserting this relation to the definition of $F'$ we now find that
new function $F$ that depends on $\bj$ and hence the action has the form 
\begin{eqnarray}
&&	S_{BMIG}=
	M^2_{BI}M^2_p\int d^4x [\sqrt{-\det \bA}+j^\alpha\partial_\alpha\phi \sqrt{-g}-\lambda\sqrt{-g}F(\bj)] \ , \nonumber \\ 
	\nonumber \\
\end{eqnarray}
that has the same form as action studied in previous section when we take into account that $-\lambda$ can be absorbed into $F$.
In other words the naive generalization of BIMG proposed in (\ref{SBIC}) is equivalent to the Born-Infeld inspired gravity minimally coupled to current after series of redefinition of non-dynamical metric and current.

\section{Hamiltonian Formalism}\label{third}
In this section we perform canonical analysis of BIMG  coupled to the perfect fluid as was introduced in the section (\ref{second}).  Since the gravitational part is the same as in standard BMIG  whose Hamiltonian
 formulation was discussed recently in \cite{Kluson:2025uyl} we write immediately its canonical form  
\begin{eqnarray}\label{actfinal}
	&&	S_{BMIG}=\int d^4x (\partial_t h^{ij}\pi_{ij}+p_\phi\partial_t\phi+\Omega \tilde{\mC}+\Omega^i\tilde{\mC}_i+\nonumber \\
	&&	+M^2_{BI}  
	M_p^2\sqrt{h}(-\frac{N^2}{\Omega}+\frac{1}{\Omega}(\Omega^i+N^i)m_{ij}(\Omega^j+N^j)-
	\Omega h^{ij}m_{ij})-\nonumber \\
	&&-M^2_{BI}M^2_p N\sqrt{m} \ ,
	\nonumber \\
\end{eqnarray}
where
\begin{eqnarray}
	&&	\tilde{\mC}=-4 M^2_{BI}M^2_p\sqrt{h}+\frac{1}{M_p^2\sqrt{h}}(\pi^{ij}h_{im}h_{jn}\pi^{mn}-\frac{1}{2}
	\pi^2)-M_p^2\sqrt{h}{}^{(3)}R \ , \nonumber \\
	&&	{}^{(3)}R= h^{ij}(-\partial_i\gamma^p_{pj}+
	\partial_m\gamma^m_{ij}-
	\gamma^n_{mi}\gamma^m_{nj}+
	\gamma^m_{ij}\gamma^p_{pm}) \ , \quad 
	\tilde{\mC}_i=-2D_k (h_{ij}\pi^{jk})\ , \nonumber \\
\end{eqnarray}
where $h_{ij}, i,j=1,2,3$ is three dimensional dynamical metric with conjugate momenta $\pi^{ij}$ and $D_k$ is three dimensional covariant derivative compatible with the metric $h_{ij}$
\begin{equation}
	D_k h_{ij}=0
\end{equation}
so that three dimensional Christoffell symbols $\gamma^k_{ij}$ are defined in a standard way
\begin{equation}
	\gamma^k_{ij}=\frac{1}{2}h^{km}(\partial_i h_{mj}+
	\partial_j h_{mi}-\partial_m h_{ij}) \ . 
\end{equation}
Further, $\Omega$ and $\Omega^i$ are Lagrance multipliers for constraints $\tilde{\mC},\tilde{\mC}_i$ respectively. Note also that $N,N^i$ and $m_{ij}$ are components of the metric $g_{\mu\nu}$ where  we used $3+1$ decomposition of metric $g_{\mu\nu}$
\cite{Dirac:1958sc,Arnowitt:1962hi,Gourgoulhon:2007ue} when we introduced   the lapse
function $N=1/\sqrt{-g^{00}}$ and the shift function
$N^i=-g^{0i}/g^{00}$. In terms of these variables we
write  components of the metric $g_{\mu\nu}$ as
\begin{eqnarray}
	g_{00}=-N^2+N_i m^{ij}N_j \ , \quad g_{0i}=N_i \ , \quad
	g_{ij}=h_{ij} \ ,
	\nonumber \\
	g^{00}=-\frac{1}{N^2} \ , \quad g^{0i}=\frac{N^i}{N^2} \
	, \quad g^{ij}=m^{ij}-\frac{N^i N^j}{N^2} \ .
	\nonumber \\
\end{eqnarray}
In case of the canonical analysis of the current contribution we proceed in the similar way as in \cite{Brown:1992kc}. As it is clear from the structure of the action (\ref{act1}) the momenta conjugate to the components of current $j^\mu$ are zero while momentum conjugate to $\phi$ is equal to
\begin{equation}\label{pphi}
	p_\phi=\frac{\partial \mL}{\partial (\partial_t \phi)}=
\sqrt{-g}j^0 
\end{equation}
so that canonical form of the action is
\begin{eqnarray}
	S_{matt}=
\int d^4x (p_\phi\partial_0\phi+p_\mu\partial_0 j^\mu-\mH)=\nonumber \\
\int d^4x (p_\phi \partial_0\phi+N\sqrt{m}j^i\partial_i\phi+\sqrt{m}N
F(\bj)) \ . \nonumber \\
\end{eqnarray}
In the standard treatment of systems of constraints we should 
interpret an absence of momenta conjugate to $j^\mu$ as primary constraints and then check the stability of them during the time evolution of system \cite{Dirac}. However this procedure is equivalent to solving equations of motion for non-dynamical variables and putting them into an action
\cite{Faddeev:1988qp,Jackiw:1993in} which we now follow. Explicitly, using (\ref{pphi}) we can write $\bj^2$ to be equal to
\begin{eqnarray}
\bj^2=-j^\mu g_{\mu\nu}j^\nu
=\frac{1}{m}p_\phi^2-
\left(\frac{p_\phi}{\sqrt{m}N}N^i+j^i\right)m_{ij}\left(
\frac{p_\phi}{\sqrt{m}}N^j+j^j\right) \ . 
\end{eqnarray}
Then we introduce independent variable $t^i$ as
\begin{equation}
	t^i=\frac{p_\phi}{\sqrt{m}N}N^i+j^i
\end{equation}
so that the action has the form 
\begin{equation}
S_{matt}=\int d^4x (p_\phi\partial_t\phi+\sqrt{m}N t^i\partial_i\phi-N^ip_\phi \partial_i\phi+N\sqrt{m}F(p_\phi,t)) \ . 
\end{equation}
Since $t^i$ are independent variables without its own dynamics it is natural to integrate it out from the action by solving corresponding equations of motion
\begin{equation}\label{partialphi}
\partial_i\phi-\frac{dF}{d\bj}\frac{1}{\bj}m_{ij}t^j=0 \ , 
\end{equation}
where we now presume that these equations can be solved for $t^i$. Note also that (\ref{partialphi}) implies
\begin{equation}
	t^i\partial_i\phi=\frac{dF}{d\bj}\frac{1}{\bj}\frac{p_\phi^2}{m}
	-\frac{dF}{d\bj}\bj
\end{equation}
and hence we obtain final form of the canonical action for current
\begin{eqnarray}
&&	S_{matt}=
\int d^4x \left(p_\phi\partial_t\phi+N\left(\frac{p_\phi^2}{\sqrt{m}}
\frac{dF}{d\bj}\bj+\sqrt{m}(F-\frac{dF}{d\bj}\bj)\right)\right)	=\nonumber \\
&&=\int d^4x (p_\phi\partial_t\phi-N\mH_T-N^i\mH_i) \ , \nonumber \\
&&\mH_T=(\frac{p_\phi^2}{\sqrt{m}}
(\rho+p)-\sqrt{m}p) \ , \quad \mH_i=p_\phi\partial_i\phi  \ 	\nonumber \\
\end{eqnarray}
that has similar form as the canonical action derived in 
\cite{Brown:1992kc}.

An important property of the whole action $S=S_{BIMG}+S_{matt}$ is that still contains non-dynamical fields $N,N^i$ and $m_{ij}$. We will eliminate them from the action by solving their equations of motion and inserting the result back to $S$, following procedure
performed in \cite{Kluson:2025uyl} so that we can immediately write the resulting action 
\begin{eqnarray}
	&&	S
	=\int d^4x 
	(\partial_t h^{ij}\pi_{ij}+p_\phi\partial_t\phi+\Omega^i(\tilde{\mC}_i+\mH_i)\nonumber \\
	&&+\Omega
	(\tilde{\mC}-\frac{1}{4M^2_{BI}M^2_p\sqrt{h}}
	\mH_i m^{ij}\mH_j
	+\frac{\lambda}{2}\frac{\sqrt{m}}{\sqrt{h}}\mH_T\nonumber \\
	&&	+M^2_{BI}M^2_p\frac{\lambda^2}{4}\frac{m}{\sqrt{h}}
	-\frac{1}{4M^2_{BI}M^2_p\sqrt{h}}\mH_T^2+M^2_{BI}M^2_p
	\sqrt{h}h^{ij}m_{ij}) \ , \nonumber \\
	\nonumber \\
\end{eqnarray}
where $m_{ij}$ should be determined from the equation
\begin{eqnarray}
	&&	-\sqrt{h}h^{ij}+\frac{1}{4}\lambda^2 \frac{\det m}{\sqrt{h}}
	m^{ij}+
	\frac{1}{4M^4_{BI}M^4_{p}\sqrt{h}}m^{ik}\mH_k\mH_l m^{lj}+
	\nonumber \\
	&&	+	\frac{1}{4}\frac{\mH_T}{M^2_p M^2_{BI}\sqrt{h}}\lambda 
	\sqrt{m}m^{{ij}}
	+\frac{1}{2M^2_{BI}M^2_p}(\lambda\sqrt{m}{\sqrt{h}}+
	\frac{\mH_T}{M^p_2M_{BI}^2\sqrt{h}})\frac{\delta \mH_T}{
		\delta m^{ij}}=0 \ .
	\nonumber \\
\end{eqnarray}
This equation should be solved for $m_{ij}$ at least in principle. 
 Further, the canonical form of the 
action is sum of the four terms proportional to Lagrange multipliers $\Omega,\Omega^i$ so that it has standard form of 
the action for diffeomorphism invariant system.

{\bf Acknowledgement:}

This work  is supported by the grant “Dualitites and higher order derivatives” (GA23-06498S) from the Czech Science Foundation (GACR).

\end{document}